\documentclass[10pt,conference]{IEEEtran}
\IEEEoverridecommandlockouts

\usepackage[nolist,nohyperlinks]{acronym}
\usepackage[backend=bibtex, style=ieee]{biblatex}
\usepackage{amsmath}
\usepackage{amssymb}
\usepackage{mathtools}
\usepackage{todonotes}
\usepackage{caption}
\usepackage[labelformat=simple]{subcaption}

\usepackage{hyperref}

\begin{acronym}

\acro{1D}{one-dimensional}

\acro{2D}{two-dimensional}

\acro{16-QAM}{16 - quadrature amplitude modulation}

\acro{64-QAM}{64 - quadrature amplitude modulation}

\acro{3GPP}{3rd.~generation partnership project}

\acro{3G}{third generation}

\acro{4G}{fourth generation}

\acro{5G}{fifth generation}

\acro{6G}{sixth generation}

\acro{AD}[A/D]{analog-to-digital}

\acro{ADSL}{asynchronous digital subscriber line}

\acro{AGC}{automatic gain control}

\acro{AM}{amplitude modulation}

\acro{AP}{access point}

\acro{ASIC}{application-specific integrated circuit}

\acro{AWGN}{additive white gaussian noise}

\acro{BB}{base-band}

\acro{BC}{broadcast channel}

\acro{BD}{block diagonalization}

\acro{BER}{bit error rate}

\acro{BoI}{bandwidth of influence}

\acro{BPSK}{binary phase shift keying}

\acro{BS}{base station}

\acro{CDF}{cumulative distribution function}

\acro{CDMA}{code division multiple access}

\acro{CN}{core network}

\acro{CNN}{convolutional neural networks}

\acro{CQI}{channel quality indicator}

\acro{CR}{cognitive radio}

\acro{CSI}{channel state information}

\acro{CSIR}{channel state information at the receiver}

\acro{CSIT}{channel state information at the transmitter}

\acro{CSMA/CA}{carrier sense multiple access with collision avoidance}

\acro{DA}[D/A]{digital-to-analog}

\acro{DFT}{discrete Fourier transform}

\acro{DL}{deep learning}

\acro{DoF}{degrees of freedom}

\acro{DPC}{dirty paper coding}

\acro{DSA}{dynamic spectrum access}

\acro{DSP}{digital signal processor}

\acro{EDGE}{enhanced data rates for \acs{GSM} evolution}

\acro{EE}{energy efficiency}

\acro{EMI}{electromagnetic interference}

\acro{EVM}{error vector magnitude}

\acro{FDD}{frequency division duplexing}

\acro{FFT}{fast Fourier transform}

\acro{FM}{frequency modulation}

\acro{FPGA}{field-programmable gate array}

\acro{FR1}{frequency range 1}

\acro{FR2}{frequency range 2}

\acro{FR3}{frequency range 3}

\acro{FSS}{frequency selective surfaces}

\acro{FTTH}{fiber-to-the-home}

\acro{GSM}{global system for mobile communications}

\acro{GPP}{general purpose processor}

\acro{HF}{high frequency}

\acro{HSPA}{high speed packet access}

\acro{IA}{interference alignment}

\acro{IC}{index coding}

\acro{IDFT}{inverse discrete Fourier transform}

\acro{IEEE}{Institute of Electrical and Electronics Engineers}

\acro{IF}{intermediate frequency}

\acro{iFFT}{inverse fast Fourier transform}

\acro{i.i.d.}{independent and identically distributed}

\acro{IoT}{internet of things}

\acro{IRBD}{iterative regularized block diagonalization}

\acro{ISM}{industrial, scientific and medical}

\acro{KKT}{Karush-Kuhn-Tucker}

\acro{LAN}{local area network}

\acro{LRMC}{low rank matrix completion}

\acro{LoS}{line of sight}

\acro{LS}{least squares}

\acro{LTE}{Long Term Evolution}

\acro{MAC}{multiple access channel}

\acro{MCS}{modulation and coding scheme}

\acro{MF}{matched filter}

\acro{MIMO}{multiple input - multiple output}

\acro{MMSE}{minimum mean square error}

\acro{MU-VFDM}{multi-user Vandermonde-subspace frequency division multiplexing}

\acro{MVNO}{mobile virtual network operator}

\acro{MUE}{macro-cell user equipment}

\acro{NC}{network coding}

\acro{OFDM}{orthogonal frequency division multiplexing}

\acro{OFDMA}{orthogonal frequency division multiple access}

\acro{ORBF}{opportunistic random beamforming}

\acro{PAN}{personal area network}

\acro{PAPR}{peak to average power ratio}

\acro{PC}{personal computer}

\acro{PCS}{personal communications systems}

\acro{PHY}{physical layer}

\acro{PLL}{phase-locked loop}

\acro{PMR}{private mobile radio}

\acro{PoC}{proof of concept}

\acro{QoS}{quality of service}

\acro{QPSK}{quadrature phase shift keying}

\acro{R-CNN}{recursive-\ac{CNN}}

\acro{RAN}{radio access network}

\acro{RAT}{radio access technology}
\acroplural{RAT}[RATs]{radio access technologies}

\acro{RB}{resource block}

\acro{RF}{radio frequency}

\acro{RIBF}{regularized inverse beamforming}

\acro{RIS}{reconfigurable intelligent surface}
\acroplural{RIS}{reconfigurable intelligent surfaces}

\acro{RX}{receiver}

\acro{SE}{spectral efficiency}

\acro{SDR}{software defined radio}

\acro{SIC}{successive interference cancellation}

\acro{SINR}{signal to interference plus noise ratio}

\acro{SIR}{signal to interference ratio}

\acro{SISO}{single input - single output}

\acro{SMMSE}{successive minimum mean square error}

\acro{SNR}{signal to noise ratio}

\acro{SR}{software radio}

\acro{SUE}{small-cell user equipment}

\acro{SUS-ZFBF}{semi-orthogonal user selection \ac{ZFBF}}

\acro{SVD}{singular value decomposition}

\acro{TDD}{time division duplexing}

\acro{TDMA}{time division multiple access}

\acro{TCP/IP}{transmission control protocol / internet protocol}

\acro{TIM}{topological interference management}

\acro{TX}{transmitter}

\acro{UE}{user equipment}

\acro{USB}{universal serial bus}

\acro{USRP}{universal software radio peripheral}

\acro{UWB}{ultra wide band}

\acro{VFDM}{Vandermonde-subspace frequency division multiplexing}

\acro{WIFI}{wireless fidelity}

\acro{YOLO}{You Only Look Once}

\acro{ZF}{zero forcing}

\acro{ZFBF}{zero force beamforming}

\end{acronym}

\bibliography{ris}

\begin{document}
\setuptodonotes{inline}
\title{An Experimental Assessment of the Spatial and Frequency Selectivity of Reconfigurable Intelligent Surfaces \thanks{This work was supported by the European Commission under Horizon Europe SNS project INSTINCT (grant 101139161), and by the French ANR (grant ANR-23-CHR4-0001-01) under the CHIST-ERA project CHASER (CHIST-ERA-22-WAI-01). Measurements presented in this work were carried out using \cxlb, a precursor infrastructure of SLICES-RI (see \url{http://wiki.cortexlab.fr} and \url{https://www.slices-ri.eu}). The RIS device was kindly provided by Greenerwave.}
}

\author{\IEEEauthorblockN{Cyrille Morin, Leonardo S. Cardoso,\\ Maxime Guillaud}
\IEEEauthorblockA{INSA Lyon, Inria, CITI Laboratory (UR3720) -- Villeurbanne, France}
\and
\IEEEauthorblockN{Ahmad Shokair, Youssef Nasser, \\ Amelie Hennequart, Geoffroy Lerosey\\ }
\IEEEauthorblockA{Greenerwave -- Paris, France}}

\newcommand{\cxlb}[0]{CorteXlab}

\maketitle

\begin{abstract}
This work investigates the impact of reconfigurable intelligent surfaces (RIS) on radio links other than the one for which the RIS configuration is optimized. We consider three different scenarios in which a secondary communication link could be affected by a RIS whose configuration is optimized for a primary communication link operating in the vicinity, on the same or on different frequencies. This question is investigated experimentally in the FR1 band, using the CorteXlab radio testbed and a Greenerwave RIS. We show that the impact, in terms of received power and impact on the channel phase of the secondary link, is significant even outside of the nominal frequency range of the RIS, and is not mitigated by carrier frequency separation between the two communication links.
\end{abstract}


\section{Introduction}
\acresetall
\Ac{RIS} is one of the key technologies touted for the \ac{6G} of cellular communications, due to its capability to improve coverage by actively overcoming shadowed areas through reflected beam steering to and from a \ac{UE} and a \ac{BS}. 
It is accomplished by modifying the \ac{RIS}'s array of reflector elements reflectivity pattern ~\cite{liang2019large, wu2021intelligent}. 
This reflectivity pattern is typically chosen to optimize a performance metric relevant to network operation.

Many works have concentrated on the positive impacts of \ac{RIS}, such as~\cite{liang2019large, wu2021intelligent, pan2021reconfigurable} and references therein.
A considerably smaller body of work deals with the potential drawbacks of \ac{RIS}. 
The work in \cite{alexandropoulos2023ris} introduces the problem of operators coexistence when a \ac{RIS} is involved; it discusses the \ac{BoI} of a single \ac{RIS} element.
More recently, \cite{pradhan2024ris}, tackled the coexistence problem consisting in mitigating the spurious reflections from a \ac{RIS} in a non-cooperating scenario, while \cite{Khateeb2026} evaluates the impact of uncoordinated RIS in the uplink on the interference between different networks.
\cite{zhao2022coexistence} introduced an analytical model for coexistence between two networks subject to a \ac{RIS} and proposed two solutions: a multilayered \ac{RIS} with out-of-band filtering capabilities and a \ac{RIS} radio blocking system that relies on angle separation to perform \ac{RIS}-enabled interference cancellation to secondary receivers.
In~\cite{dejesus2022interference}, the authors have studied the effects of surrounding and uncontrolled \ac{EMI} on ongoing communications using a \ac{RIS} to enhance communications \ac{SNR}. 
\Ac{FSS}  \cite{Chen_etal_frequency_selective_surface_survey_CommsSurveyTuto24} are another possible solution to the coexistence issue; by selectively transmitting or reflecting electromagnetic waves in a specific frequency range, they can help limit the secondary wideband effect of \ac{RIS}.

In this work, we aim to contribute to the analysis of the potentially deleterious effects of \ac{RIS} from an experimental point-of-view by measuring its effects of co-channel and adjacent channel communications. 
Unlike~\cite{dejesus2022interference,zhao2022coexistence,pradhan2024ris}, we provide actual radio measurements to support an insightful and comprehensive analysis of coexistence in three cases: 1) the impact when an operator deploys a \ac{RIS} device, to other nearby users of the same operator (operating on the same frequency band), 2) the potential impact on nearby users of coexisting operators, and 3) the impact of a \ac{RIS} optimized for one band to signals transmitted to the same user on another band. 

This paper is organized as follows. 
Section~\ref{sec:impact} details the cases studied in which the adoption of a \ac{RIS} device can be nefarious to nearby users. 
We present our measurement scenario in section~\ref{sec:measurements}. 
A comprehensive analysis of the measurements, explaining our metrics is given in section~\ref{sec:analysis}. 
We present and discuss the findings in section~\ref{sec:results}, and finally draw conclusions and discuss possible solutions in section~\ref{sec:conclusion}.

\maketitle

\section{\acs{RIS} Impact on Communications}\label{sec:impact}

\begin{figure*}[h]
    \centering
    \begin{subfigure}{0.35\textwidth}
        \centering
        \includegraphics[trim=0 75 0 10,clip,width=0.99\textwidth]{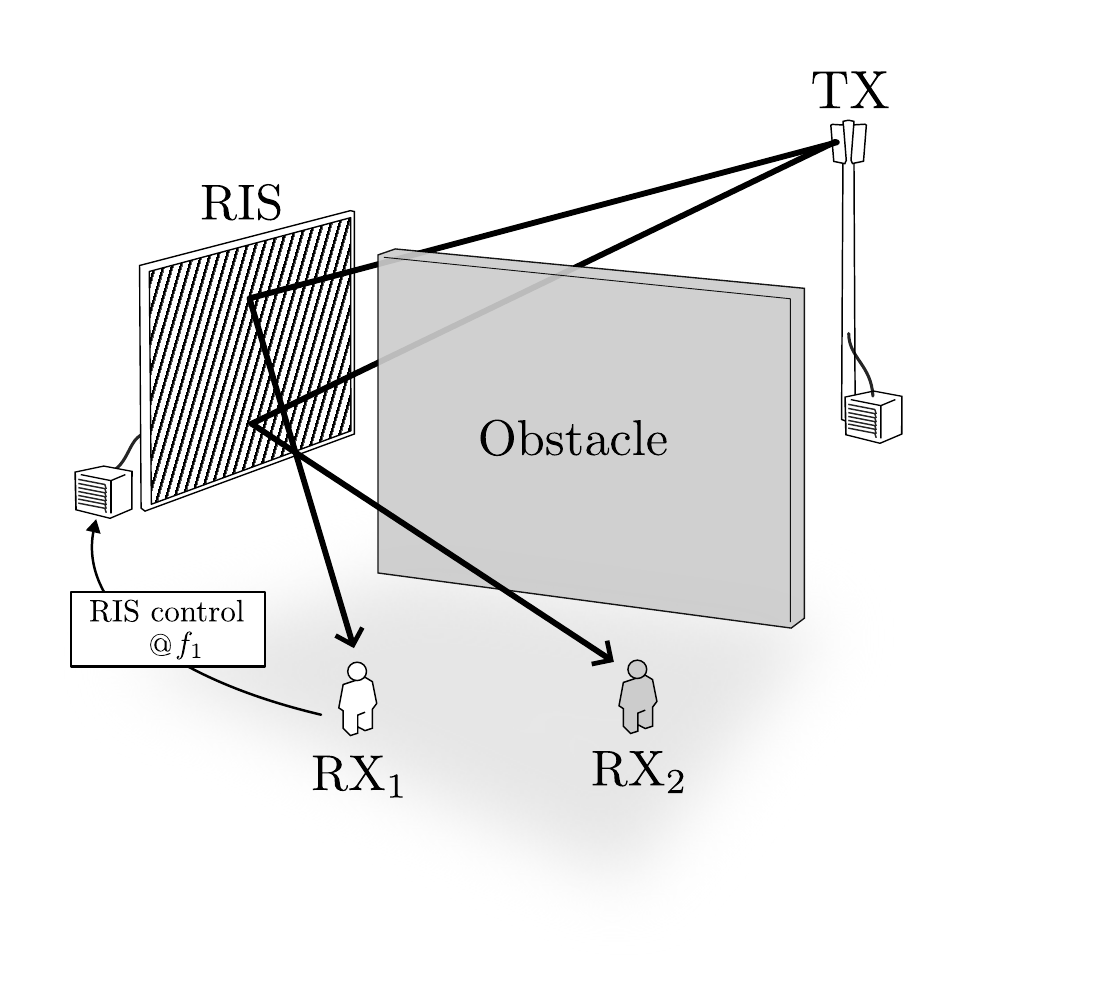}
        \caption{TDMA scenario.}
        \label{fig:scenario_tdma}
    \end{subfigure}\hspace{-0.6cm}
    \begin{subfigure}{0.35\textwidth}
        \centering
        \includegraphics[trim=0 75 0 10,clip,width=0.99\textwidth]{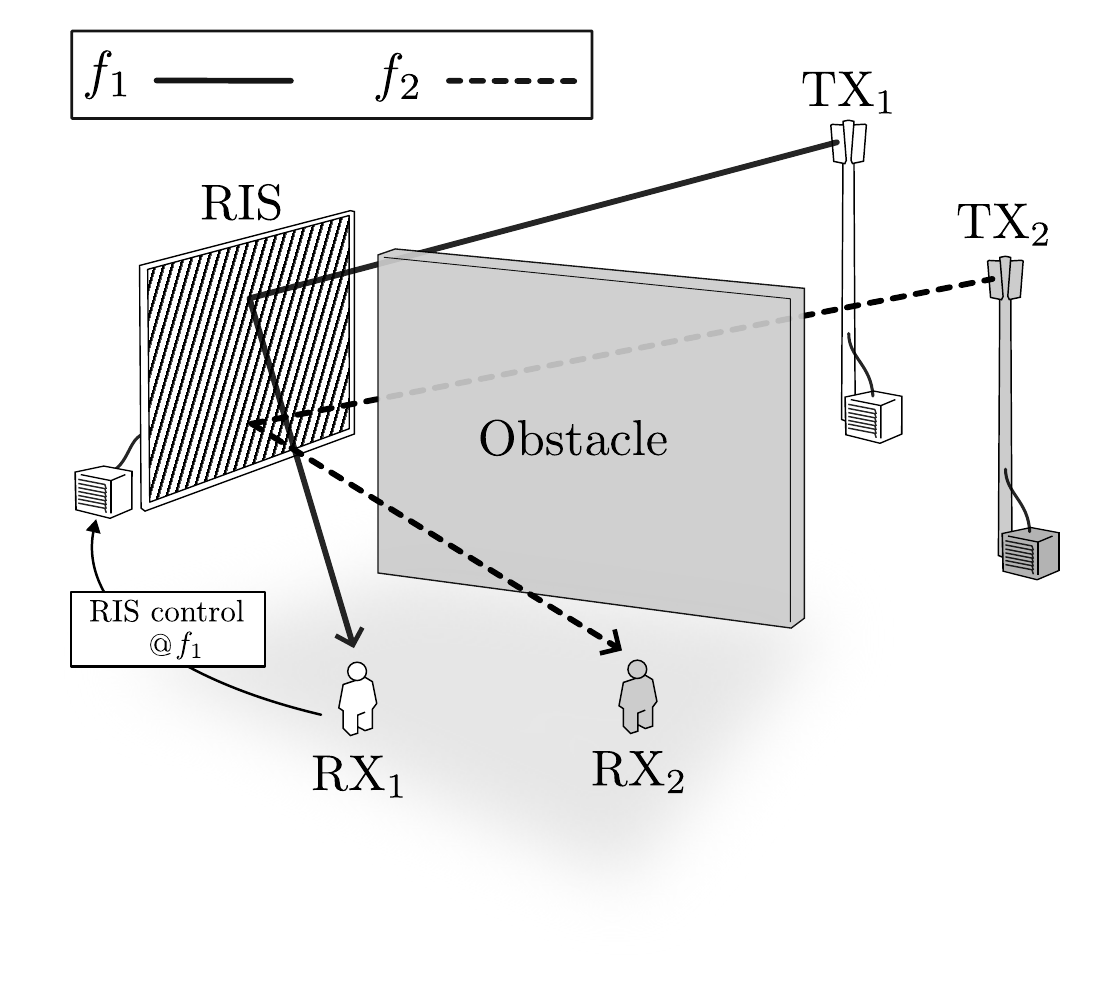}
        \caption{Multiple Operators Scenario.}
        \label{fig:scenario_multi-op}
    \end{subfigure}\hspace{-0.6cm}
    \begin{subfigure}{0.35\textwidth}
        \centering
        \includegraphics[trim=0 75 0 10,clip,width=0.99\textwidth]{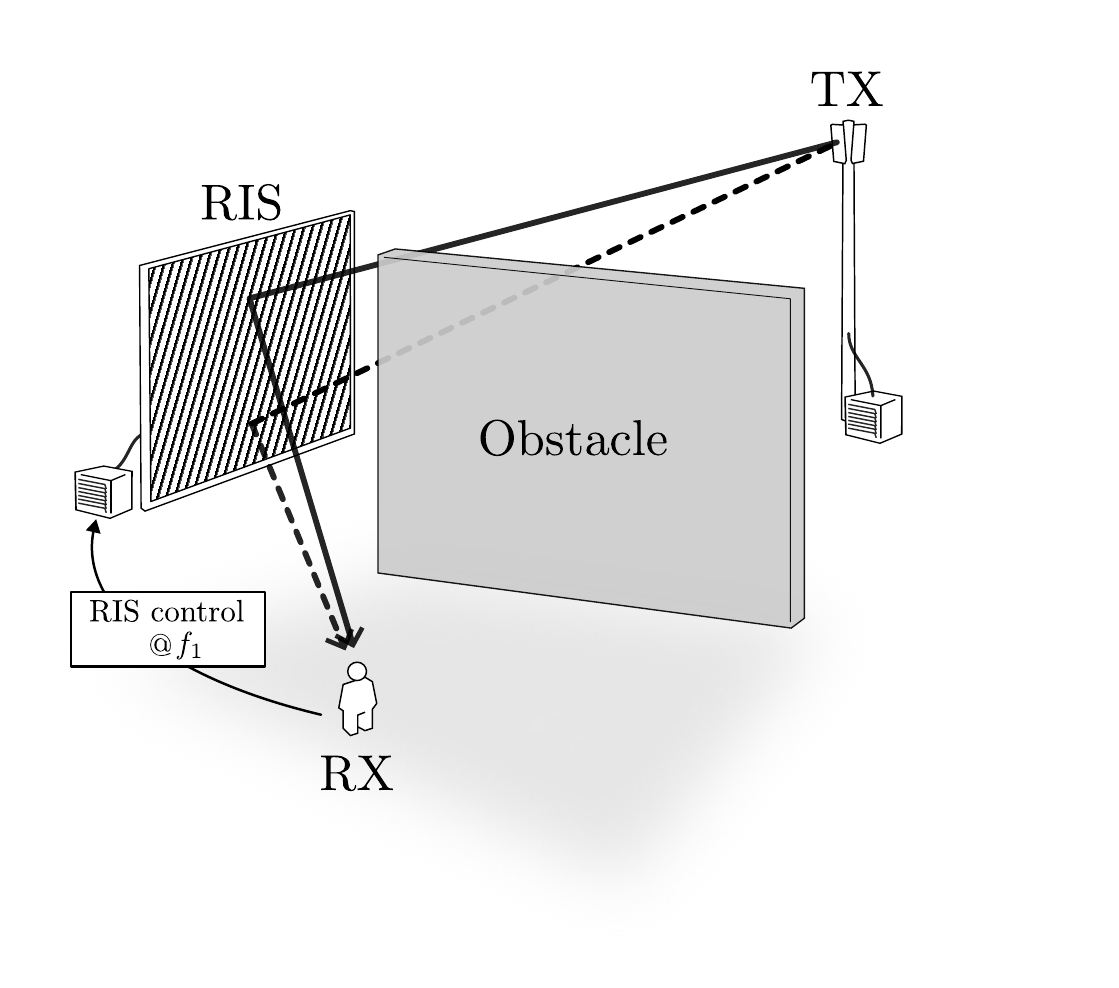}
        \caption{Multiband Scenario.}
        \label{fig:scenario_multi-band}
    \end{subfigure}
    \caption{Scenarios in which a \ac{RIS} can negatively impact the communication performance of the secondary link}
    \label{fig_scenarios}
\end{figure*}

In this work, we will consider a \emph{primary} link, between \ac{TX}$_1$ and \ac{RX}$_1$ operating at frequency $f_1$, for which the \ac{RIS} parameters are being optimized, as well as a \emph{secondary} communication link, between \acs{TX}$_2$ and \acs{RX}$_2$, operating in parallel to the primary link.
The assumption is that the secondary system is affected by the \ac{RIS}, however not necessarily in the most favorable way since the \ac{RIS} is optimized for the primary system.
In fact, in many deployment cases, the primary and secondary system may not even be aware of the presence of each other, for instance in the case of two cellular operators using adjacent carriers.
We want to understand the consequences of the \ac{RIS} usage from the primary link over the secondary link, both in terms of power gain/loss (a metric commonly considered to evaluate positive RIS effects in the literature) and in terms of the sudden changes in channel phase due to changes in the \ac{RIS} configuration.
Changes in channel phase may have little or no impact on the received power, but can be harmful to coherent transmission schemes or beamforming strategies.
These metrics are detailed in Section~\ref{sec:analysis}.
Our study is structured around three representative scenarios, which are detailed next.


\subsection{TDMA Scenario}\label{sec_tdma_scenario}

The scenario involves the primary and secondary communication links sharing a same frequency ($f_1=f_2$), with \acs{TX}$_1$ and \acs{TX}$_2$ co-located, as depicted in Figure~\ref{fig:scenario_tdma}. 
This may happen in the case where two or more users of the same network operator, communicating with the same base station, are scheduled alternatively over the same frequency (or resource block, as in modern \ac{OFDM}-based networks with dynamic resource block allocation).
We assume that the \ac{RIS} is optimized for \acs{RX}$_1$ only (for instance due to the limited overhead dedicated to \ac{RIS} control), even when the intended user is \acs{RX}$_2$. 
We hence evaluate the impact on \acs{RX}$_2$ of a \ac{RIS} configuration which has been optimized for another user in the network, operating on the same frequency. 


\subsection{Multi-Operators Scenario}\label{sec_multi_operators_scenario}

This scenario corresponds to the case where two networks operate independently on distinct frequency bands ($f_1 \neq f_2$) in a shared geographical area, as depicted in~Figure~\ref{fig:scenario_multi-op}.
Such a scenario is often encountered when different operators are allocated distinct frequency ranges in the same frequency band, with typical separation between center frequencies ranging from dozens to hundreds of MHz.
We assume that the \ac{RIS} is optimized for the primary link, and that there is no coordination with the secondary operator.
We evaluate the impact of the \ac{RIS} controlled by first operator to optimize the  \acs{TX}$_1$-\acs{RX}$_1$ link on the operation of the \acs{TX}$_2$-\acs{RX}$_2$ link.
Note that it is sometimes assumed that the effect of a \ac{RIS} can be limited to a band of interest, and that its impact on adjacent bands will be negligible; however we will see that this assumption is disproven by our experiments.



\subsection{Multiband Scenario}\label{sec_multiband_scenario}

The last scenario, depicted in Figure~\ref{fig:scenario_multi-band}, involves a single transmitter and a single receiver, however the communication is performed on two distinct frequency bands,  with primary and secondary links on frequencies $f_1$  and $f_2$ respectively; the \ac{RIS} is configured to enhance the quality of the link on frequency $f_1$ only.
Note that the lack of channel reciprocity over distinct frequencies implies that optimizing on frequency $f_1$ is not necessarily favorable for frequency $f_2$.
This case may arise due to \ac{FDD} uplink-downlink switching, or in systems using multi-band carrier aggregation or frequency hopping.
Hence, in this scenario, we evaluate the impact of using a \ac{RIS} optimized for the correct \ac{TX} and \ac{RX} locations but for an incorrect frequency.


For all of the above three scenarios, solutions to mitigate the deleterious effect of the RIS on the secondary system exist and can be implemented with varying degrees of technical complexity. In the TDMA and multiband scenarios, mitigation through \ac{RIS} re-configuration is possible since the (single) \ac{TX} is aware of the radio parameters of all links (primary and secondary); however there may be cases where this reconfiguration is not desirable or impossible, e.g. to limit signaling overhead or because \ac{RX}$_2$ is not in a connected state. 
For the multi-operator scenario, the situation is more complicated since coordination between several operators, potentially using hardware from different vendors, would be technically complex.
Thus, we believe that the proposed experimental assessment is
of interest as it can motivate the necessity of policies to guide coordination in the use of time, frequency and spatial resources. 
This study may also highlight problems concerning the usage of autonomous \ac{RIS} devices~\cite{croisfelt2025autonomous}.

\section{Measurement Protocol}\label{sec:measurements}

\cxlb ~is a physical-layer-centric radio testbed composed of 40 \ac{SDR} nodes, composed of a mix of \ac{SISO} and \ac{MIMO} nodes, with synchronization infrastructure to provide time and frequency synchronization across radios and hosted in an electromagnetically isolated room of approximately 160~m$^2$. It allows for stable radio propagation and reproducible experiments since it does not suffer interference from the outside. Table~\ref{tab:cxlb} provides the important features of \cxlb.
\begin{table}[htb]
    \centering
    \begin{tabular}{c|c}
       \hline
       Room dimensions  & 18~m $\times$ 9~m\\
       \hline
       ~ & no reflections on walls\\
       Propagation details & only reflections off floor,\\
       ~ & and hardware on ceiling\\
       \hline
       Antennas & log periodic AARONIA 70600\\
       \hline
       Antenna grid layout  & every 1.8~m, in two dimensions\\
       \hline
       Antenna height from floor & 1.65~m\\
         \hline
    \end{tabular}
    \vspace{1em}
    \caption{\cxlb~characteristics}
    \label{tab:cxlb}
\end{table}
Measurements were carried out inside of \cxlb~using a \ac{RIS}
device targeting the \ac{FR1} band \cite{Greenerewave_RIS_FR1_specsheet,}.
Its main characteristics are summarized in table~\ref{tab:ris}.
\begin{table}[htb]
    \centering
    \begin{tabular}{c|c}
      \hline
      Operating frequency range & 3.3-4.1GHz (\ac{FR1})\\
      \hline
      Panel dimensions   & 40~cm $\times$ 40~cm\\
      \hline
      Number of elements  & 128\\
      \hline
      Number of states per element   & 2 (on-off)\\
      \hline
      Polarization & H-pol and V-pol \\
      \hline
      Scan Range (Azimuth, Elevation) & +/- 60°\\
      \hline
      RF Power Dissipation at Reflection & max 1.5dB, average 1dB\\
      \hline
      RF Power Handling & 50W\\
      \hline
      Average Power Consumption & $\leq$ 0.2W\\
      \hline
    \end{tabular}
    \vspace{1em}
    \caption{Greenerwave FR1 \ac{RIS} device characteristics}
    \label{tab:ris}
\end{table}

\begin{figure}[h]
    \centering
    \includegraphics[trim=0 0 0 102,clip,width=0.89\columnwidth]{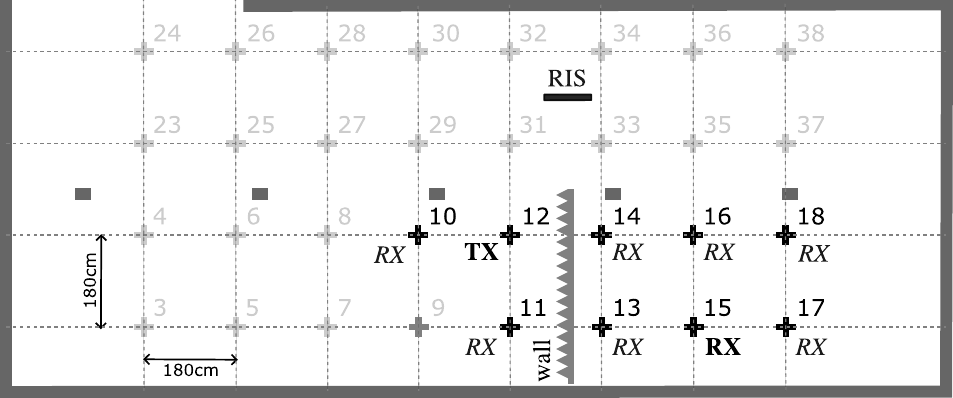}
    \caption{Layout of measurements in \cxlb. Crosses indicate antenna positions for the respective nodes. Dark grey indicates room structure and light grey the unused nodes.}
    \label{fig:layout_cortexlab}
\end{figure}

To reproduce a configuration where using a \ac{RIS} would be relevant, a wall covered with aluminum and RF absorbing foam panels is placed in the room, blocking \ac{LoS} communication between groups of nodes 10-11-12 and 13-14-15-16-17.
The \ac{RIS} is installed in-line with this wall so as to be in \ac{LoS} of both groups of nodes (see Fig.~\ref{fig:layout_cortexlab}).
A \ac{TX} is active on one side of the wall (node 12), with most of the \acp{RX} (nodes 13 to 18) shadowed by the wall.
To provide comparison points, two \acp{RX} are also set next to the \ac{TX} (nodes 10 and 11).

The channel sounding system operates as follows:
All radio devices are synchronized in time and frequency using 10MHz and PPS signals from an Octoclock device, ensuring a common carrier frequency and sampling time and rate.
The active transmitter emits a continuously repeating sequence obtained as the inverse \ac{DFT} of the Zadoff-Chu sequence of length 255 and root index 1
at a rate of $5.10^6$ samples per second.
The received sequences are integrated four times, to increase \ac{SNR} and reduce computation loads, before going through a \ac{DFT}.
Symbol-wise division by the original sequence yields the frequency-domain channel responses, sampled in 255 bins spanning 5 MHz (19.6~kHz subcarrier spacing).
Measurements proceed in two phases, namely (i) RIS configuration and (ii) measurements collection.

During phase (i), the \ac{RIS} configuration is optimized to maximize the average power (in dB) for the primary link (\ac{TX}$_1$-\ac{RX}$_1$), measured in the middle 1 MHz of the transmission band and excluding the bin at DC (0~Hz).
A codebook of \ac{RIS} configurations is constructed by optimizing the \ac{RIS} parameters using the algorithm described in \cite{chen2024RISoptimization} for different values of $f_1$ and for different target receivers; the objective is to be able to randomize the \ac{RIS} configuration in a way that ensures that all states in the codebook correspond to a reasonable operating state for the primary link (it is optimal for some choice of a \ac{RX}$_1$ node and frequency $f_1$).
For the TDMA scenario, the 1901-element codebook is constructed by letting $f_1$ span frequencies ranging from 2.5~GHz to 4.4~GHz in 1~MHz steps and is restricted to $RX_1=15$, 
while for the Multi-Operator and Multi-band scenarios, the 768-element codebook spans the same frequency band in 20~MHz steps, for $RX_1$ spanning nodes 10, 11 and 13 to 18.
During phase (ii), another sweep is performed, over the same frequency range and step size for $f_2$.
At each step, all configurations in the codebook are recalled successively, and for each, the entire complex frequency response is recorded at all receivers, along with a reference measurement with the \ac{RIS} in its default (off) state.

\section{Analysis}\label{sec:analysis}


\begin{figure*}[h]
    \centering
    
        
    
    \begin{subfigure}{0.95\textwidth}
        \centering
        \includegraphics[trim=0 0 0 0,clip,width=\textwidth]{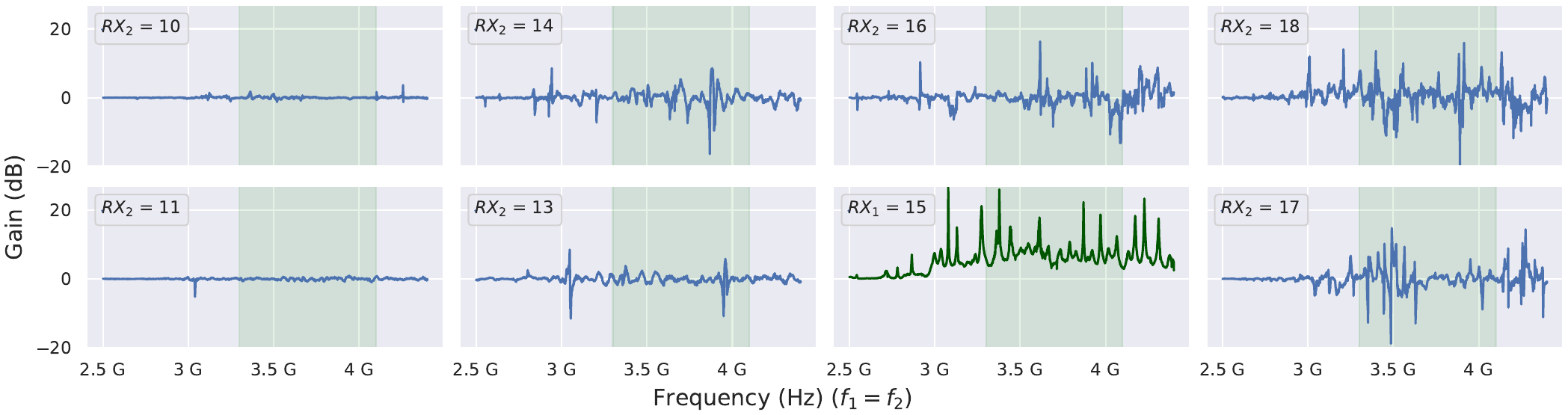}
        \caption{Gains}
        \label{fig:scenario1_Gain}
    \end{subfigure}

    \begin{subfigure}{0.95\textwidth}
        \centering
        \includegraphics[trim=0 0 0 0,clip,width=\textwidth]{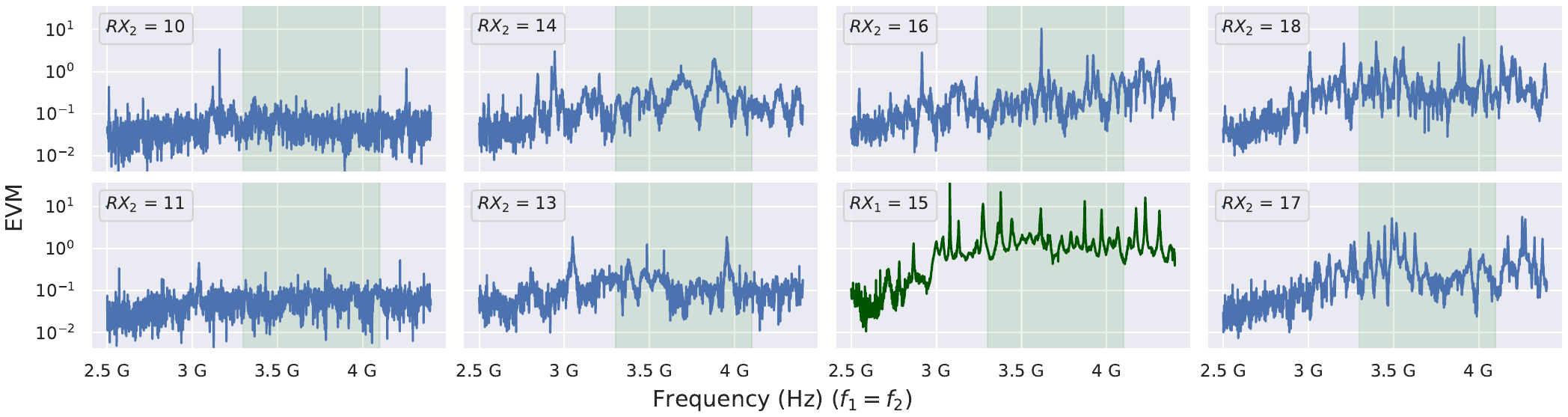}
        \caption{EVM}
        \label{fig:scenario1_EVM}
    \end{subfigure}
    
    \caption{TDMA Scenario, $f_1 = f_2$, $1$~MHz steps. For all plots, RX$_1$ = 15. Green shade is nominal RIS operation band}
    \label{fig:scenario1_results}
\end{figure*}

\begin{figure*}[h]
    \centering
    \begin{subfigure}{0.95\textwidth}
        \centering
        \includegraphics[trim=0 0 0 21,clip,width=0.99\textwidth]{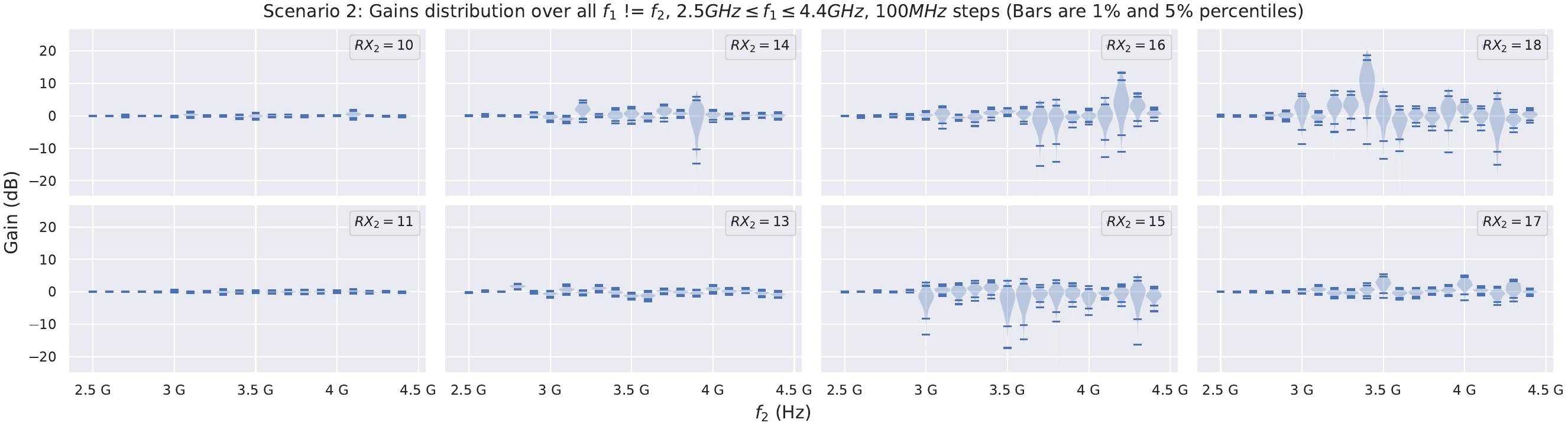}
        \caption{Gains}
        \label{fig:scenario2_Gain}
    \end{subfigure}

    \begin{subfigure}{0.95\textwidth}
        \centering
        \includegraphics[trim=0 0 0 21,clip,width=0.99\textwidth]{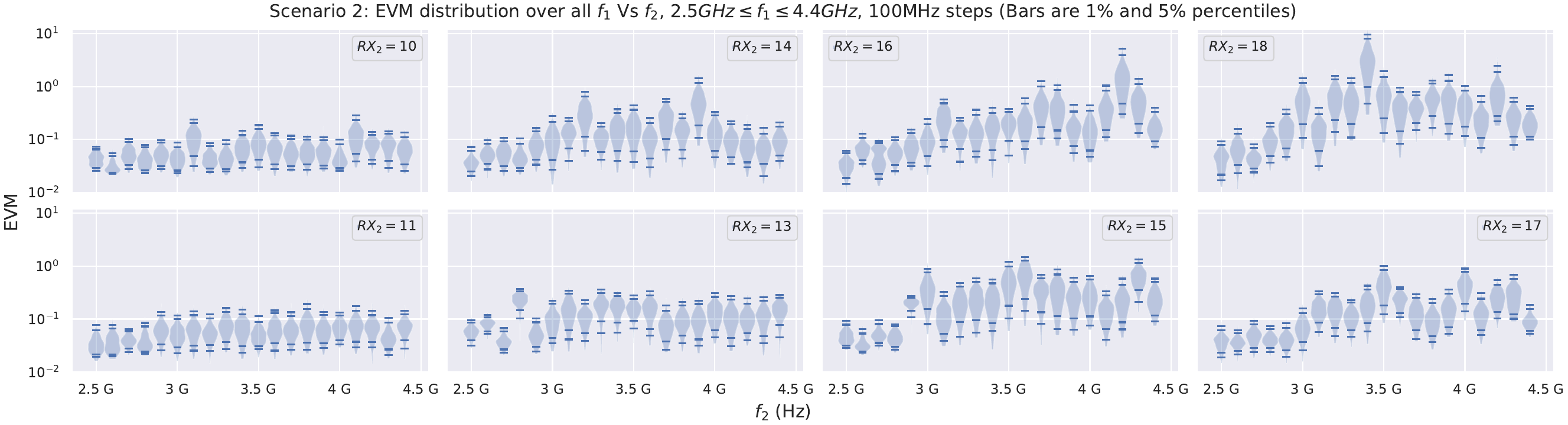}
        \caption{EVM}
        \label{fig:scenario2_EVM}
    \end{subfigure}
    
    \caption{Multiple Operators Scenario, distribution over all $RX_1 \neq RX_2$ and all $f_1 \neq f_2$, $2.5~\text{GHz} \leq f_1 \leq 4.4~\text{GHz}$, in $100$~MHz steps (bars denote 1\% and 5\% percentiles).}
    \label{fig:scenario2_results}
\end{figure*}

The gathered dataset contains channel frequency response vectors for all the combinations of $f_1$, $f_2$, $RX_1$, and $RX_2$.
Two metrics are extracted from this data, and interpreted according to the three different scenarios described in Section~\ref{sec:impact}.

\subsubsection{TX$_2$-RX$_2$ Power gain/loss at frequency $f_2$}
Power at \ac{RX}$_2$ is averaged as described above over a 1~MHz band centered on $f_2$, and compared with the corresponding power of the reference (RIS off) measurement at the same $f_2$, measuring the \ac{RIS} impact on the received power.

\subsubsection{\acs{EVM} due to mismatched equalization}
This models a situation where where the \ac{RIS} configuration changes during the transmission of a data frame.
We assume a worst-case scenario where the channel gain $\mathbf{h} \in \mathbb{C}$ on the secondary link is estimated from a pilot sequence (e.g. demodulation reference symbols, DMRS), immediately followed by a \ac{RIS} re-configuration so that the subsequent data symbols experience a different channel $\mathbf{h}' \neq \mathbf{h}$.
At high \ac{SNR}, a zero-forcing equalizer would multiply the received baseband samples with $\mathbf{h}^{-1}$ to produce data symbol estimates, resulting in a mismatched equivalent gain $\mathbf{h}^{-1}\cdot\mathbf{h}' $ which is in general different from 1 due to the \ac{RIS} reconfiguration.
We evaluate this deviation in terms of the resulting \ac{EVM}, defined as
$    \mathrm{EVM} = \left| \mathbf{h}^{-1} \cdot \mathbf{h}' -1 \right|$.
For both the Gain and \ac{EVM}, metrics are computed for each frequency bin and averaged over a 1~MHz band centered on $f_2$.

\section{Results}\label{sec:results}


\subsection{\acs{TDMA} Scenario}

The first set of results corresponds to the \acs{TDMA} scenario ($f_1=f_2$) described in Section~\ref{sec_tdma_scenario} with \ac{RX}$_1$ chosen as node 15.
Power gain and \ac{EVM} metrics are presented in Figs.~\ref{fig:scenario1_Gain} and \ref{fig:scenario1_EVM} respectively. 
Subplots positions for different choices of \ac{RX}$_2$ are arranged to correspond approximately to the real spatial distribution of \acp{RX} shown in Figure~\ref{fig:layout_cortexlab}.
Looking at the subplot for \ac{RX}$_1=15$ in Fig.~\ref{fig:scenario1_Gain}, the primary user benefits from more than 5~dB of gain across the band starting from $\sim$3~GHz and up to the highest measured frequency, and reaches up to 20~dB at specific frequencies.
This validates the expected behavior of the \ac{RIS}, which was designed to operate between the 3.3 and 4.1~GHz (RIS's nominal frequency range, depicted by a green shade on the figures) and has no effect for frequencies below $\sim$3~GHz.
The frequencies where a high gain (i.e., above 10~dB) is achieved correspond to frequencies where the user would suffer from deep fading without the \ac{RIS}. We should note, however, that the \ac{RIS} provides consistent gains all over the target frequency range. As for \ac{EVM}, as seen in Fig.~\ref{fig:scenario1_EVM}, we see that the primary user's channel suffers from a high phase mismatch throughout the intended frequency range of the \ac{RIS}. This is seen for \ac{EVM} values averaging around 1 in the range, and peaking above 10 for some frequencies. This behavior is not harmful to the primary user's communications since it can estimate the channel and use proper equalization to correct for these distortions. This high \ac{EVM} is expected since the optimization algorithm the \ac{RIS} uses focuses on gain rather than channel response.

For the other \ac{RX}$_2$, two cases can be distinguished. Nodes 10 and 11 are within direct LoS of the transmitter. As expected, these two \acp{RX} see minimal gain and \ac{EVM} impact from the \ac{RIS}, especially since the \ac{RIS} is optimized for a node on the other side of the separating wall. The \ac{EVM} in particular remains quite low, with the exception of a few peaks above 1 at certain frequencies.
On the other hand, when \ac{RX}$_2$ is one of nodes 13-18 (for which the \ac{TX}-\ac{RX}$_2$ link is blocked by the wall), with the \ac{RIS} still configured to optimize the primary link between \ac{TX} and  \ac{RX}$_1$ (node 15), the effect is more pronounced, with positive gains up to 15 and negative gains up to -20~dB, depending on the frequency. These results highlight a frequency selectivity induced by the \ac{RIS} for masked nodes. As for the \ac{EVM} (Fig.~\ref{fig:scenario1_EVM}), significant channel perturbation is incurred by the secondary link, with different behaviors depending on the secondary \ac{RX} position. 

\subsection{Multiple Operators Scenario}\label{sec:res_multi_operators}

The second set of results corresponds to the scenario described in Section~\ref{sec_multi_operators_scenario}, where the effect on the secondary link is measured for each choice of $RX_2$ and $f_2$, where $RX_1 \neq RX_2$ and $f_1 \neq f_2$.
For simplicity, the measurements are restricted to co-located $TX_1$ and $TX_2$ (Node 12).
Results are depicted on Figs.~\ref{fig:scenario2_Gain} (gains) and \ref{fig:scenario2_EVM}  (EVM).
As in the previous scenario, the two secondary \acp{RX} with \ac{LoS} to the transmitter, i.e. nodes 10 and 11, both see minimal impact due to the \ac{RIS}. When \ac{RX}$_2$ is one of nodes 13-18, the effect of the \ac{RIS} reconfiguration is noticeable for frequencies $f_2$ above 3~GHz approximately.
In most frequencies, the gain distribution is centered around 0~dB but in some, this center is offset by 5 or even 10~dB. 
These correspond to frequencies $f_2$ where the $RX_2$ suffers from deep fading in the absence of \ac{RIS}; in that situation, it is more likely that a \ac{RIS} configuration improves the received power than degrades it.
In contrast, in some cases, the distribution, though centered around 0~dB, shows a long tail towards negative gains.
At these frequencies, the channel is prone to disturbances from the \ac{RIS}.
Both cases exhibit a correspondingly high \ac{EVM}.
We note that even at frequencies where the impact in terms of power gain/loss remains low, the \ac{EVM} distributions are centered above $.1$ in most cases, which can be sufficient to create significant equalization mismatch.

\subsection{Multiband Scenario}
\begin{figure*}
    \centering
    \begin{subfigure}{0.66\textwidth}
        \centering
        \includegraphics[trim=0 0 0 22,clip,width=1\textwidth]{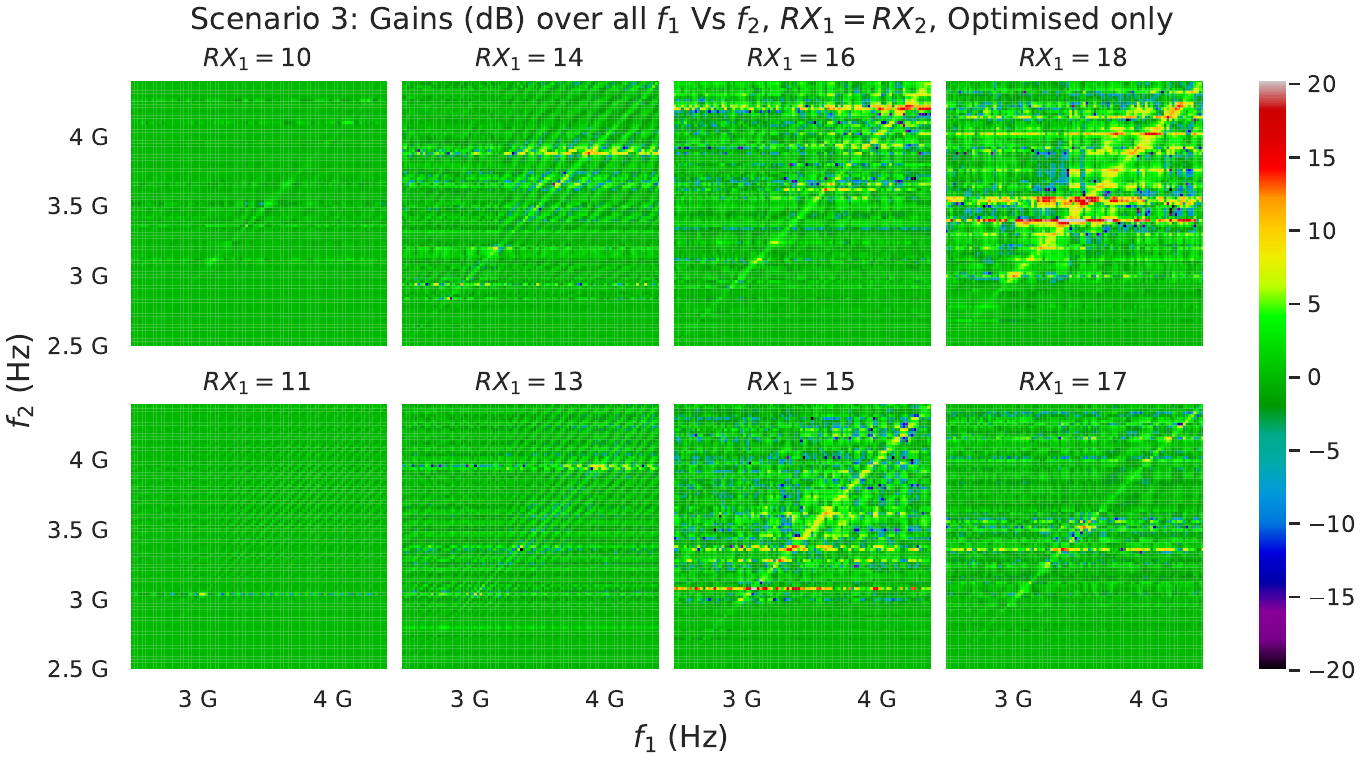}
        \caption{Gains (dB)}
        \label{fig:scenario3_Gain}
    \end{subfigure}
    \begin{subfigure}{0.66\textwidth}
        \centering
        \includegraphics[trim=0 0 0 22,clip,width=1\textwidth,]{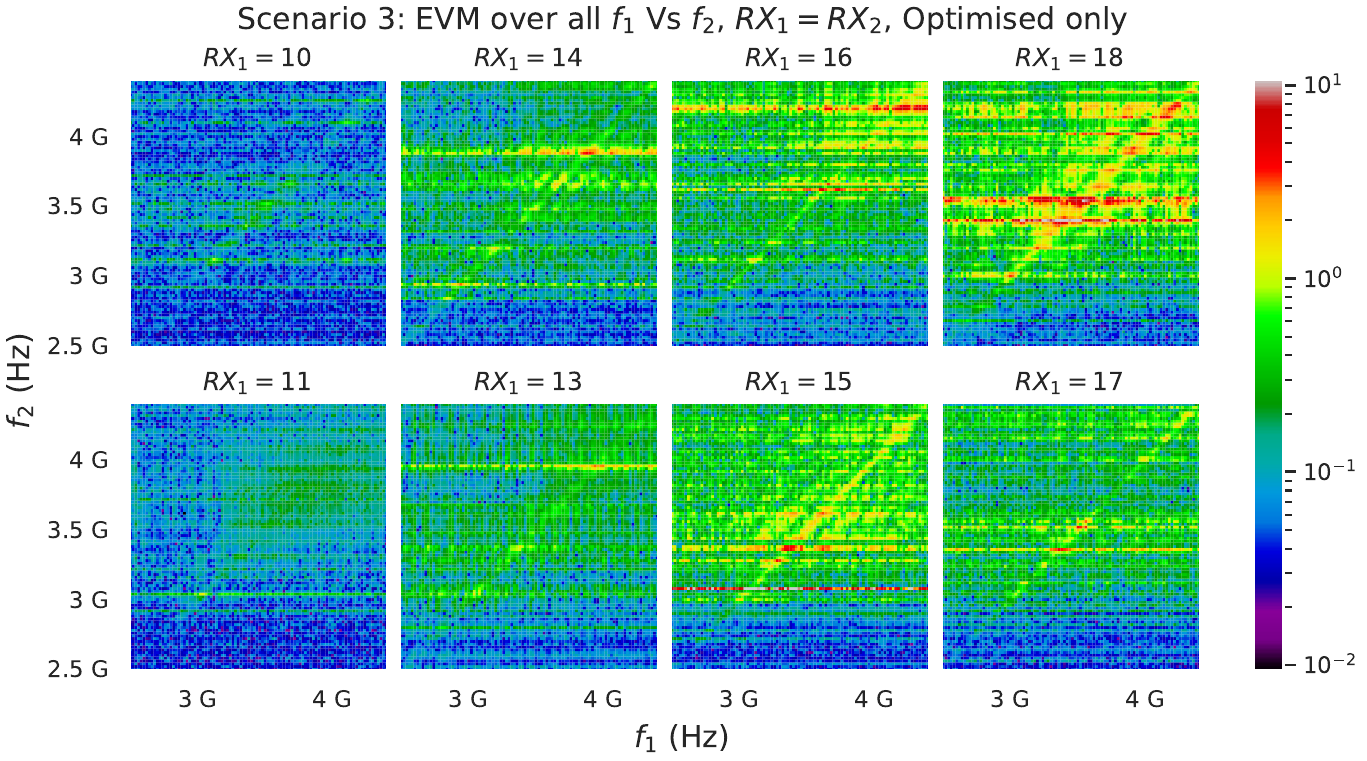}
        \caption{EVM}
        \label{fig:scenario3_EVM}
    \end{subfigure}
    
    \caption{Multiband Scenario, Gain and EVM metrics over all $f_1$ vs. $f_2$, RX$_1$ = RX$_2$}
    \label{fig:scenario3_results}
\end{figure*}
The results presented in subsection \ref{sec:res_multi_operators} are averaged for $f_1, f_2$ over the whole frequency band. 
In this section, however, we present individual data points for $(f_1, f_2)$ tuples.
The results, depicted in Fig.\ref{fig:scenario3_results}, correspond to the scenario described in Section~\ref{sec_multiband_scenario} where we consider all combinations of $f_1$ and $f_2$ for cases where $TX_1 = TX_2$ and $RX_1 = RX_2$.
The diagonal on all plots corresponds to cases where $f_1 = f_2$ (primary and secondary systems coincide), showing the intended effect of the \ac{RIS}.
Again, receivers with \ac{LoS} (i.e. nodes 10 and 11) and frequencies below $\sim 3~GHz$ incur minimal impact from the \ac{RIS}.
High gain areas, which show as horizontal lines in Figs.~\ref{fig:scenario3_Gain}, correspond to frequencies $f_2$ where the secondary link is in deep fades and is likely to benefit from a random RIS configuration. 
Diagonal ripples can be seen in both Gain and \ac{EVM} metrics around the main diagonal line, alternating between higher and lower impacts with different periods depending on the receiver between $\sim80$~MHz for RX 11 and 13, and $\sim300$~MHz for RX 18.
As in the previous scenario, the \ac{EVM} is above .1, and even above 1 in large parts of the measured spectrum, with green and yellow dominating plots for receivers without \ac{LoS} (\ac{RX}$_2 \in \{13, 14, 16, 17, 18\}$).
Even for $RX_2 = 11$ that has barely any visible power gain impact, a patch of \ac{EVM} higher than 0.2 can be seen with $f_1$ and $f_2$ around $\sim3.75$~GHz, illustrating that even channel phase shifts yielding small changes in the received power can significantly degrade transmission quality.

\section{Conclusion}\label{sec:conclusion}
We have evaluated experimentally the impact of a RIS on radio links other than the one for which the RIS configuration is optimized, in the FR1 band. Our measurements show that the impact on the secondary link, in terms of received power and EVM due to mismatched equalization, is significant even outside of the nominal frequency range of the RIS, and is not mitigated by carrier frequency separation between the primary and secondary communication links.
\newpage
\renewcommand*{\bibfont}{\small}
\printbibliography

\end{document}